# Flux pinning mechanism in $NdFeAsO_{0.82}F_{0.18}$ superconductor: Thermally activated flux flow and charge carrier mean free path fluctuation pinning

X.L. Wang[1*], S. R. Ghorbani[1,2], S. X. Dou[1*]

[1] *Institute for Superconducting and Electronic Materials, University of Wollongong, Wollongong, New South Wales 2522, Australia*

[2] *Department of Physics, Tarbiat Moallem University of Sabzevar, P.O. Box 397, Sabzevar, Iran*

Xiao-Li Shen[3], Wei Yi[3], Zheng-Cai Li[3], Zhi-An Ren[3]

[3] *National Laboratory for Superconductivity, Institute of Physics, Chinese Academy of Sciences, PO Box 603, Beijing, P. R. China, 100190*

The flux pinning mechanism of $NdO_{0.82}F_{0.18}FeAs$ superconductor made under high pressure, with a critical temperature, $T_c$, of 51 K, has been investigated in detail in this work. The field dependence of the magnetization and the temperature dependence of the magnetoresistivity were measured in fields up to 13 T. The field dependence of the critical current density, $J_c(B)$, was analyzed within the collective pinning model. A crossover field, $B_{sb}$, from the single vortex to the small vortex bundle pinning regime was observed. The temperature dependence of $B_{sb}(T)$ is in good agreement with the $\delta l$ pinning mechanism, i.e., pinning associated with fluctuations in the charge-carrier mean free path, $l$. Analysis of resistive transition broadening revealed that thermally activated flux flow is found to be responsible for the resistivity contribution in the vicinity of $T_c$. The activation energy $U_0/k_B$ is 2000 K in low fields and scales as $B^{-1/3}$ over a wide field range. Our results indicate that the $NdO_{0.82}F_{0.18}FeAs$ has stronger intrinsic pinning than Bi-2212 and also stronger than $MgB_2$ for H > 8 T.

The exciting discovery of the FeAs based new superconductor with critical temperature, $T_c$, as high as 26-55 K [1-9] has opened a new chapter in the fields of high temperature superconductivity and magnetism. The newly discovered superconductors have the general formula REOTmPn, where RE is a rare earth element, Tm a transition metal, and Pn = P or As. The As based compounds exhibit higher $T_c$ than P based ones. The $T_c$ is controlled by the size of the RE ion, by electron doping, either by F substitution for O [1] or by oxygen deficiency [4] and by hole doping in La1-xSrxOFeAs [5]. Most of work reported so far has been focusing on the underlying physics of the high temperature superconductivity in this new class of superconductors. In terms of potential applications, the upper critical field, $H_{c2}$, and the critical current density, $J_c$, are the two central topics for research on the new superconductors. The $H_{c2}$ is an intrinsic property, which has been reported to be over 65 T in $LaO_{0.9}F_{0.1}FeAs$ [7,8], 150 T in $SmO_{0.85}F_{0.15}FeAs$ [9], and 80-300 T in an $NdO_{0.82}F_{0.18}FeAs$ bulk sample with $T_c$ of 51 K [10, 11]. However, the $J_c$ is sample dependent and controlled by the flux pinning behavior. The $NdO_{0.82}F_{0.18}FeAs$ superconductors fabricated under high pressure have the following features: 1) The $J_c$ values in polycrystalline samples are reported to be as low as $10^2$-$10^4$ A/cm$^2$ [10] from 5-30 K, which is 10 times lower than what exists in individual grains, which have $J_c$ of ~$10^5$ or $10^6$A/cm$^2$ [12, 13]. 2). Despite the low $J_c$, we found that the $J_c$ in the samples dropped weakly with magnetic field at low temperatures [10]. 3) From the resistivity-temperature measurements we found that the sample can carry supercurrent, even in fields of up to 9 T at 37 K. However, the magnetization vs. field measurements showed a reversible magnetization. 4) It was found that $NdO_{0.89}F_{0.11}FeAs$ is a paramagnetic superconductor with an antiferromagnetic transition at Néel temperature $T_N$ = 11 K [14]. 5) The NdOFeAs samples made under high pressure are highly dense (close to the theoretical density) and contain little secondary phase [3,10,14], which rules out weak-links contributed from impurity phases. Due to the above mentioned features, it is very important to study the flux pinning mechanism and understand the underlying physics controlling the $J_c$ behavior in relation to magnetic field for the $NdO_{1-x}F_xFeAs$ superconductor. In this work, we report that the thermally activated flux flow is responsible for the transition broadening and the $J_c$-field dependence is controlled by the charge carrier mean free path fluctuation pinning mechanism.

A polycrystalline sample with the nominal composition $NdO_{0.82}F_{0.18}FeAs$ was prepared by a high-pressure (HP) technique. Powders of NdAs, Fe, $Fe_2O_3$, and $FeF_2$ were well mixed, pelletized, and then sealed in a boron nitride crucible and sintered at 1250 °C for 2 hours under the high pressure of 6 GPa [3]. The Rietveld refinement indicated that the sample is of single phase with only a small amount of $Nd_2O_3$ [10]. Standard four-probe resistivity measurements were carried out on a bar sample by using a physical properties measurement system (PPMS, Quantum Design) in the field range from 0 up to 13 T. Magnetic loops were also collected at various temperatures below $T_c$. The magnetic hysteresis loops were measured every 2.5 K from 5-40 K. The critical current density was calculated by using the Bean approximation, $J_c = 20\Delta m/Va(1- a/3b)$ where $a$ and $b$ are the width and the length of the sample perpendicular to the applied field, respectively, V is the sample volume, and $\Delta m$ is the height of the *M-H* hysteresis loop.

The broadening of the resistive transition in magnetic field for the cuprate layered superconductors and $MgB_2$ is interpreted in terms of a dissipation of energy caused by the motion of vortices [15-17]. The resistance in the broadened region is caused by the creep of vortices, so that the temperature dependence of the resistivity, *ρ(T),* is of the thermally activated type described by the equation $\rho(T,B) = \rho_0 \exp[-U_o/k_BT]$, where $U_0$ is the flux-flow activation energy, which can be obtained from the slope of the linear part of an Arrhenius plot, $\rho_0$ is a parameter, and $k_B$ is Boltzmann's constant The $U_0$ can be deduced only from the limited temperature interval below $T_c$ where the data from the Arrhenius plot of *ρ(T)* yields a straight line.

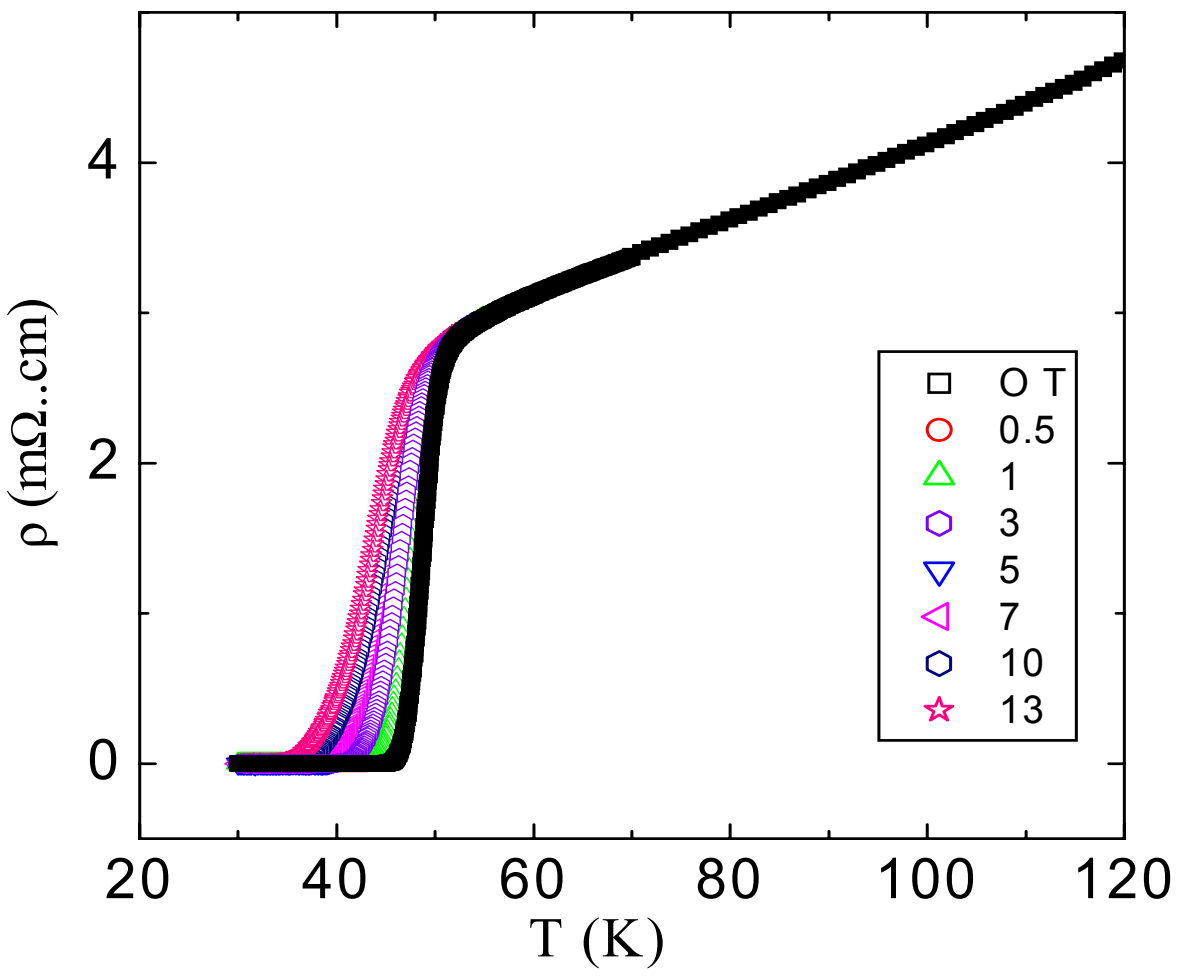

Fig. 1. Temperature dependence of an NdOFeAs sample measured in different magnetic fields up to 13 T.

Thermally activated flux flow is manifested in the broadening of superconducting transitions under various magnetic fields, as shown in Fig. 1. In Fig. 2, we plot the data as log ρ vs. $T^{-1}$. It can be seen that the thermally activated behavior of the resistance is immediately apparent. The slope of the curve is the activation energy $U_0$. The best fit to the experimental data yields values of the activation energy ranging down from $U_0/k_B = 2000$ K in the low field of 0.1 T, as shown in Fig. 2. It has been reported that for Bi-Sr-Ca-Cu-O (BSCCO) crystals, the activation energy exhibits different power-law dependences on magnetic field, $U_0(B) \sim B^{-n}$, with n = 1/2 for B < 5T and n = 1/6 for B > 5 T, in the case when field is parallel to the *c*-axis [15,17]. Figure 3 shows the magnetic field dependence (up to 13 T) of the activation energy $U_0$ of $NdO_{0.82}F_{0.18}FeAs$. We can see that the values of $U_0$ drop very weakly with field for B < 0.4 T, scaled as $B^{-0.07}$, and then decreases slowly as $B^{-1/3}$ for B > 0.4 T. The values of $U_0$ in both low and high fields are 2-3 times larger that for Bi2212 [15] and 10 times greater than for Bi2223 [17]. This indicates that the $NdO_{0.82}F_{0.18}FeAs$ has stronger intrinsic pinning than BSCCO. However, the $U_0$ for $NdO_{0.82}F_{0.18}FeAs$ is almost one order of magnitude lower than that of $MgB_2$ thin films [16] for H < 8 T, but higher than $MgB_2$ for H > 8 T. This indicates that the $NdO_{0.82}F_{0.18}FeAs$ has strong pinning characteristics in high magnetic fields, in agreement with the very weak $J_c$-field dependence that we reported previously [10] and the high values of the irreversibility field, $H_{irr}$, in an $NdO_{0.82}F_{0.18}FeAs$ single crystal [11].

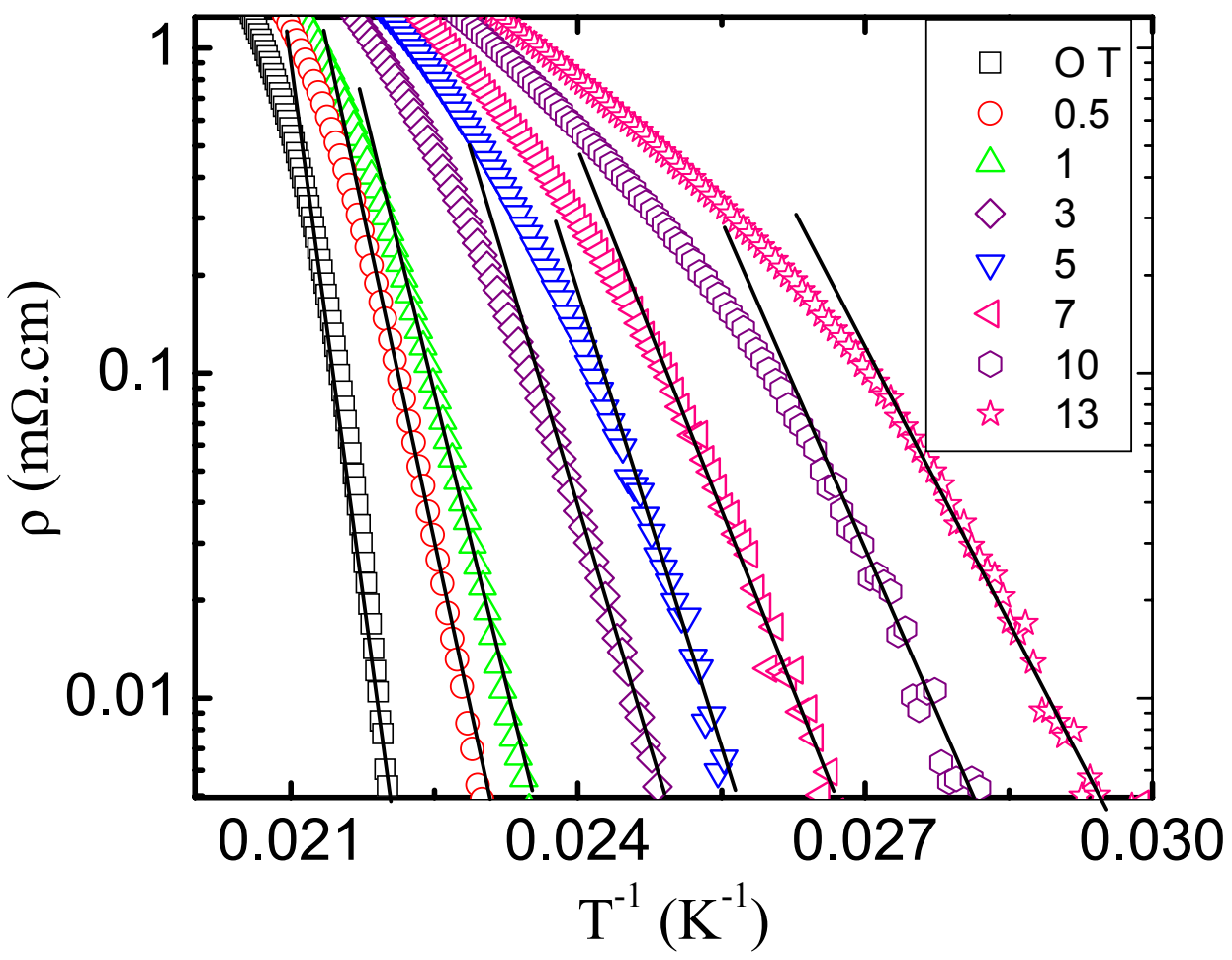

Fig. 2. Arrhenius plot of the electrical resistivity of an $NdO_{0.82}F_{0.18}FeAs$ sample**.** The activation energy $U_0$ is given by the slopes from linear fitting.

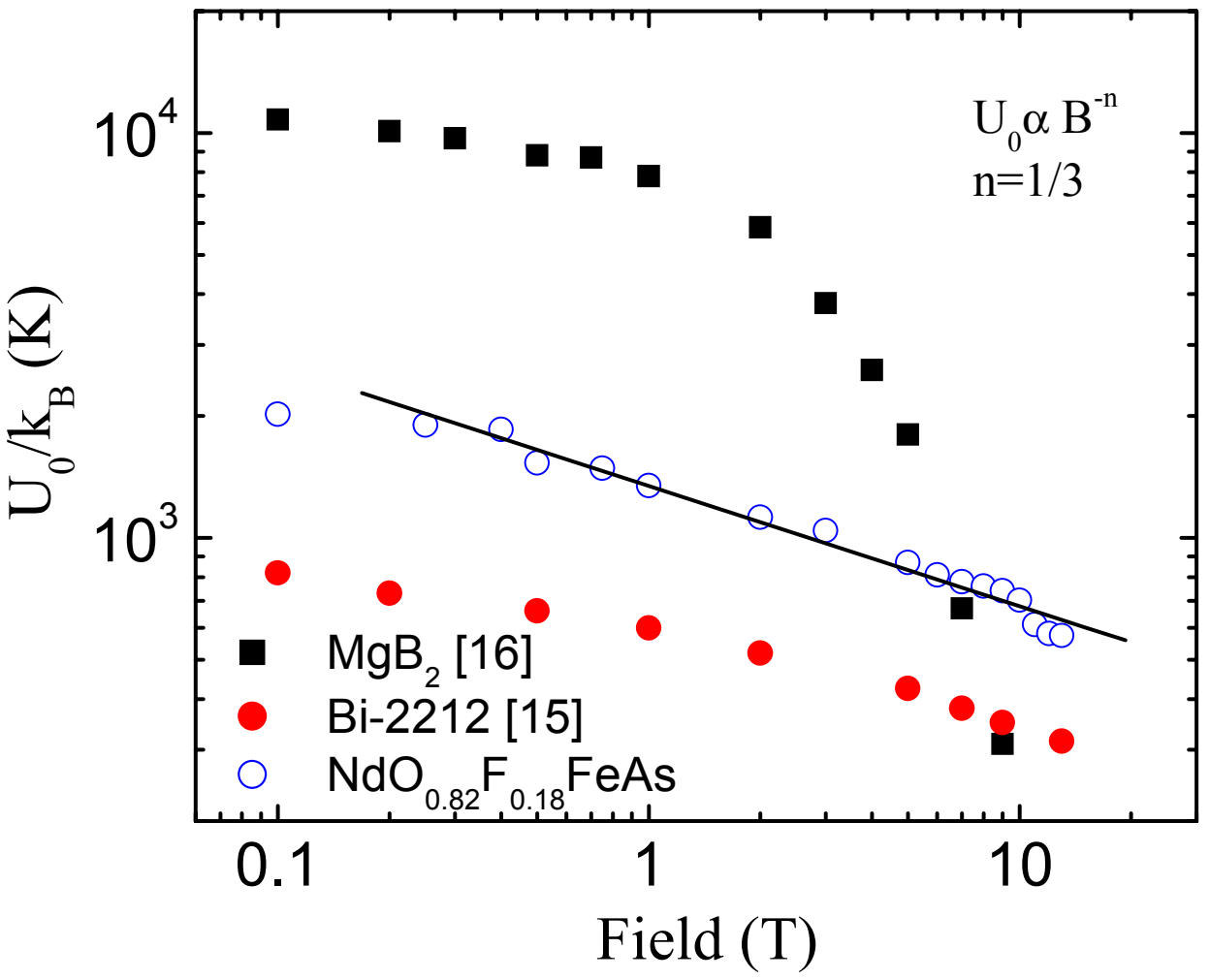

Fig. 3. Magnetic field dependence of the activation energy $U_0$ of $NdO_{0.82}F_{0.18}FeAs$. The linear portions of the data suggest the power law $U_0 \sim B^{-n}$.

The $J_c(B, T)$ results are shown in a double-logarithmic plot in Fig. 4. At 5 K, the $J_c$ value is over $1\times10^4$ A/cm$^2$. The $J_c$ initially shows a plateau at low field (H < 0.1) and then begins to decrease slowly towards saturation at H > 10 T. For T > 5 K, a second peak appears in high fields, and the peak position decreases with increasing temperature. The second peak effect has also been reported in SmOFeAs [9] and in F-free but oxygen deficient NdOFeAs samples [18].

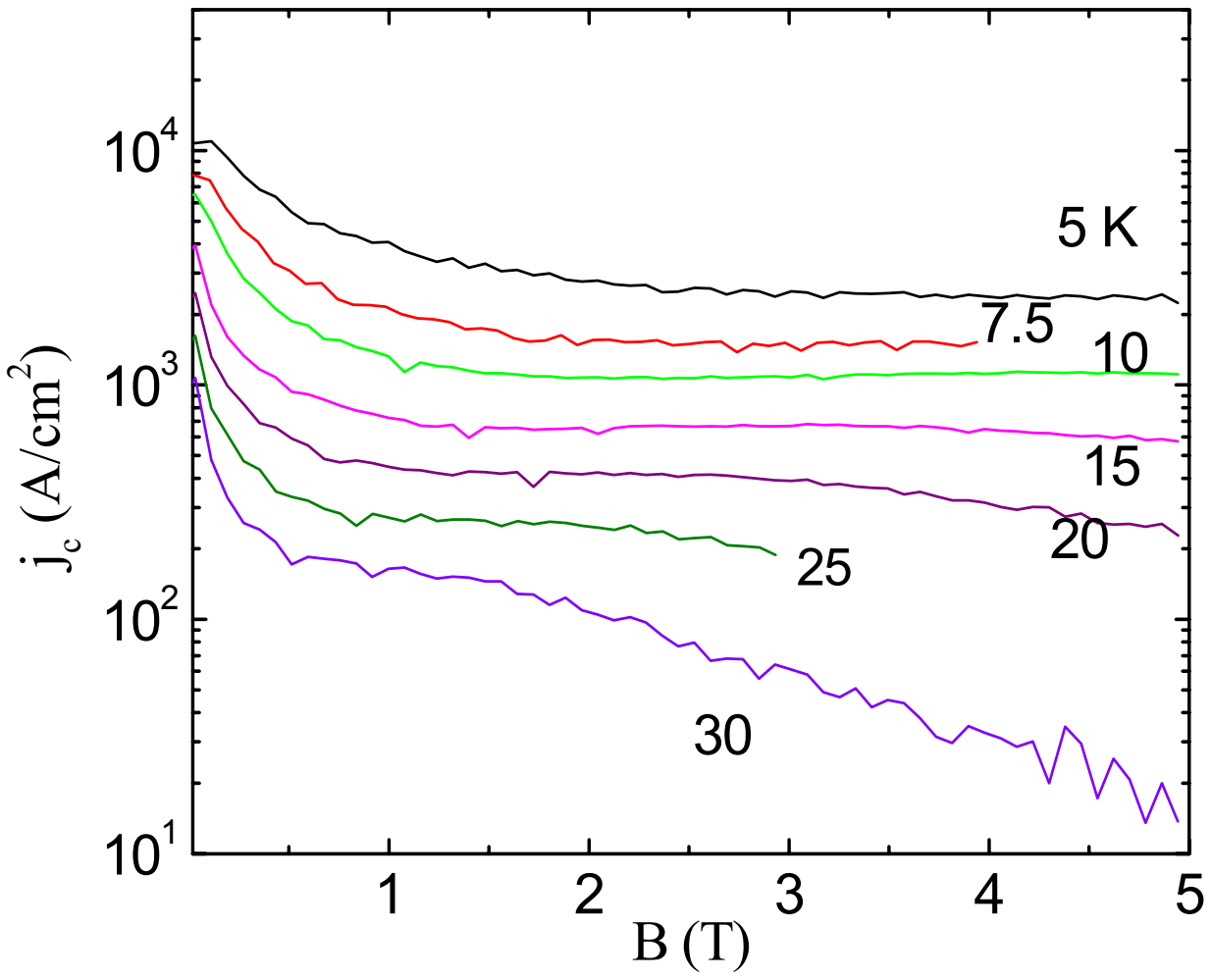

Fig. 4. The $J_c$-field dependence measured at different temperatures.

In the framework of the collective vortex pinning theory, which was derived by Blatter et al. [19], the critical current density is field independent when the applied magnetic field is lower than the crossover field, $B_{sb}$. In the regime below $B_{sb}$ a single vortex pinning mechanism governs the vortex lattice:

$$B_{sb} \propto J_{sv} B_{c2} \qquad (1)$$

where $J_{sv}$ is the critical current density in the single vortex-pinning regime. At higher fields, for $B > B_{sb}$, $J_c(B)$ decreases quickly, and it follows an exponential law:

$$J_c(B) \approx J_c(0)\exp[-(B/B_0)^{3/2}] \qquad (2)$$

where $B_0$ is a normalization parameter of the order of $B_{sb}$. For $B > B_{sb}$, a power dependence in the form of $J_c(B) \propto B^{-\beta}$ acts from $B_{sb}$ to another crossover field $B_{lb}$ (to the large bundle pinning regime).

$-\log[J_c(B)/J_c(0)]$ as a function of $B$ is shown in a double logarithmic plot in Fig. 4. As can be seen, there are two deviations from linearity, as indicated by the arrows shown in the Figure. The deviation at low fields, below $B_{sb}$, is associated with crossover from the single vortex-pinning regime to the small bundle-pinning regime. The high field deviation is related to the field at which the secondary peak effect starts to appear. The crossover field $B_{sb}$ as a function of temperature is shown in Fig. 5.

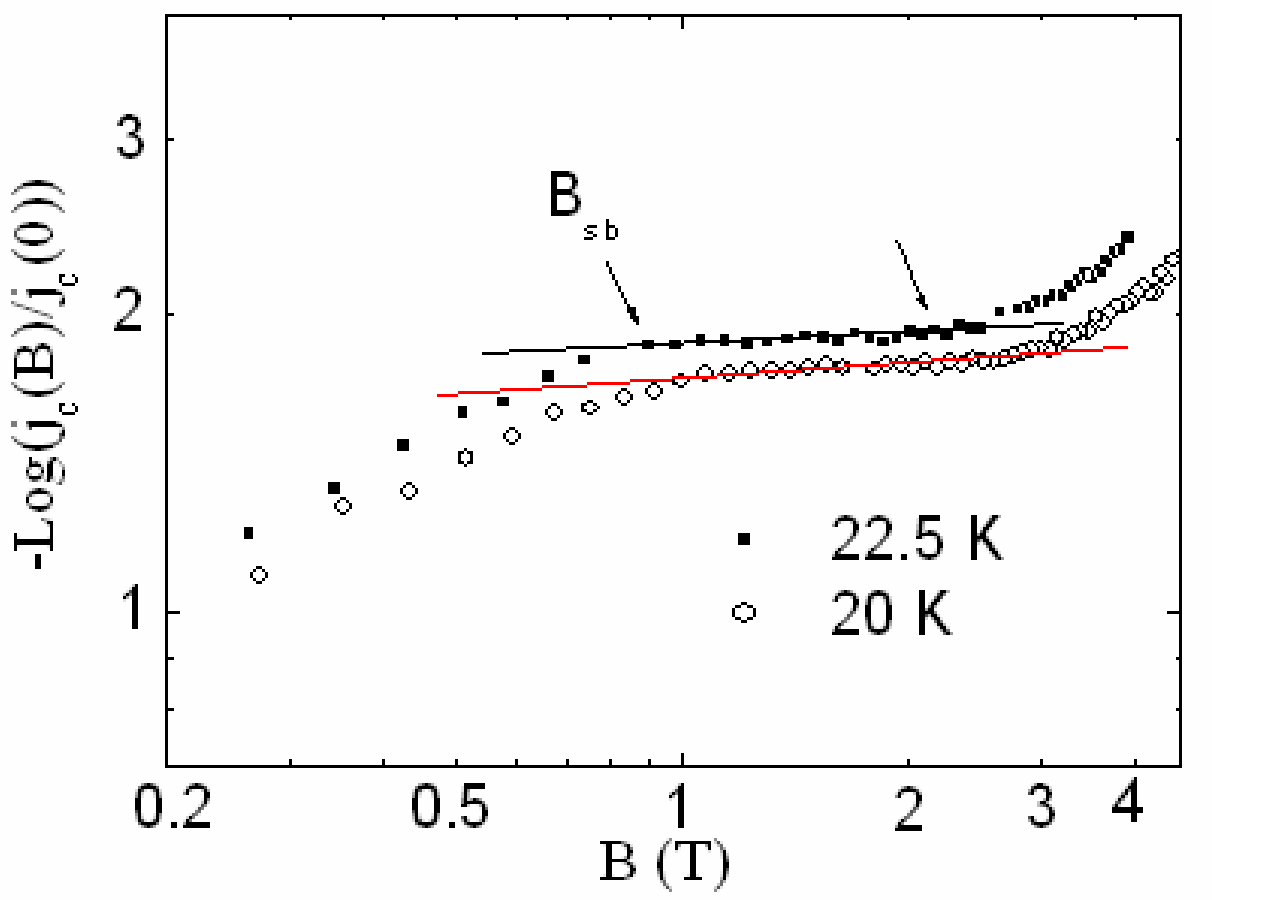

**Fig. 5**: Double logarithmic plot of $-\log[J_c(B)/J_c(0)]$ as a function of B at T = 20 and 22.5 K. The crossover fields $B_{sb}$ and *secondary peak field* are indicated by arrows.

Griessen et al [20] pointed out that the $\delta T_c$ and $\delta l$ pinning mechanisms, depending respectively on variations in the critical temperature and the mean free path, result in different temperature dependencies of the critical current density $J_{sv}$ in the single vortex pinning regime. They found that $J_{sv} \propto (1-t^2)^{7/6}(1+t^2)^{5/6}$, with $t = T/T_c$, for the case of $\delta T_c$ pinning, while for $\delta l$ pinning, $J_{sv}\ \alpha\ (1-t^2)^{5/2}(1+t^2)^{-1/2}$. Inserting these two $J_{sv}(T)$ expressions into Eq. (1), the following temperature dependence for $B_{sb}$ can be obtained

$$B_{sb}(T) = B_{sb}(0)\left(\frac{1-t^2}{1+t^2}\right)^{\upsilon} \qquad (3)$$

where υ = 2/3 and 2 for $\delta T_c$ and $\delta l$ pinning, respectively.

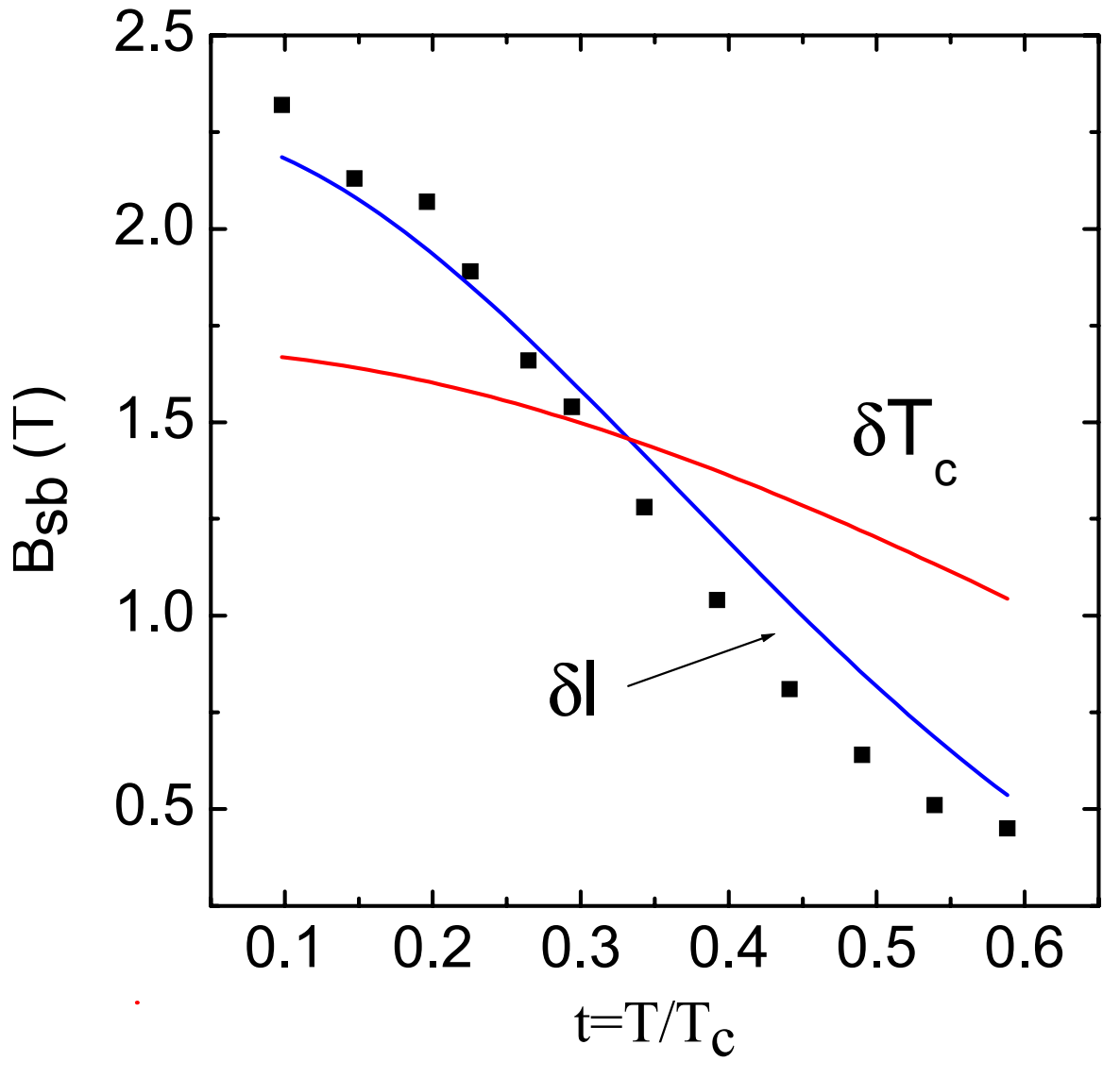

**Fig. 6**: Temperature dependence of the crossover field $B_{sb}$. The solid curves are fits to Eq. (3).

The curve has a positive curvature in the $\delta T_c$ pinning case, while the curvature associated with the $\delta l$ pinning is negative. As is clear from Fig. 6, the $B_{sb}$(T) behaviour shows a negative curvature. There is also good agreement between our experimental points and Eq. (3) with $\upsilon = 2$. This result strongly suggests that the flux pinning behavior in $NdO_{0.82}F_{0.18}FeAs$ is controlled by the $\delta l$ pinning mechanism instead of $\delta T_c$ pinning. It should be noted that the models of collective vortex pinning also applies to the $SmFeAsO_{0.9}F_{0.1}$ sample [21] which was also made under high pressure.

In summary, thermally activated flux flow is found to be responsible for the resistivity contribution in the vicinity of $T_c$ in $NdO_{0.82}F_{0.18}FeAs$ superconductor made under high pressure. The activation energy $U_0/k_B$ is 2000 K and scales as $B^{-1/3}$. $NdO_{0.82}F_{0.18}FeAs$ has stronger intrinsic pinning than Bi-2212 and also stronger than $MgB_2$ for H > 8 T. Using the collective flux pinning model, the field dependence of the magnetization shows that the flux pinning in the sample is dominated by the $\delta l$ pinning mechanism, i.e., pinning associated with charge-carrier mean free path fluctuations.

This work is supported by the Australian Research Council.

*Electronic address: xiaolin@uow.edu.au
*Electronic address: shi@uow.edu.au